\documentclass[showpacs,aps,prb,twocolumn]{revtex4}

\usepackage{graphicx}

\def  \bsig   {\mbox{\boldmath$\sigma $}}

\begin{document}

\title{Spin relaxation related to the edge scattering in graphene}

\author{V.~K.~Dugaev}
\affiliation{Department of Physics,
Rzesz\'ow University of Technology, al.~Powsta\'nc\'ow Warszawy 6,
35-959 Rzesz\'ow, Poland}
\affiliation{Departamento de F\'isica and CFIF, Instituto Superior T\'ecnico,
Universidade de Lisboa, av.~Rovisco Pais, 1049-001 Lisboa, Portugal}

\author{M. I. Katsnelson}
\affiliation{Radboud University Nijmegen, Institute for Molecules and Materials,
Heyendaalseweg 135, 6525 AJ Nijmegen, The Netherlands}
\affiliation{Department of Theoretical Physics and Applied Mathematics, Ural Federal University, 
Mira St. 19, 62002 Ekaterinburg, Russia}

\begin{abstract}
We discuss the role of spin-flip scattering of electrons from the magnetized edges in graphene nanoribbons.
The spin-flip scattering is associated with strong fluctuations of the magnetic moments at the edge.
Using the Boltzmann equation approach, which is valid for not too narrow nanoribbons, we 
calculate the spin relaxation time in the case of Berry-Mondragon and zigzag graphene edges.
We also consider the case of ballistic nanoribbons characterized by very long momentum
relaxation time in the bulk, when the main source of momentum and spin relaxation is the spin-dependent
scattering at the edges. We found that in the case of zigzag edges, an anomalous spin diffusion
is possible, which is related to very weak spin-flip scattering of electrons gliding along the nanoribbon
edge.    
\end{abstract}

\date{\today }

\maketitle


\section{Introduction}

One of the most challeging problems in the physics of graphene\cite{geim07,katsnelson} is related 
to the spin relaxation time of electrons or holes.\cite{tombros07,dlubak12} 
This is really the key point for possible spintronic applications\cite{fabian07} 
of this material.\cite{trauzettel07}
It was suggested before that the spin relaxation in graphene can be extremely weak due
to very small intrinsic spin-orbit interaction.\cite{huertas06,min06,yao07,gmitra09,pesin12}
The latter was extensively discussed in the past\cite{tombros07,tombros08,pi10,yang11,han11,mani12}
and, finally, quite persuasively confirmed by numerous investigations.\cite{dlubak12,wojtaszek13}
  
However, it is quite obvious that the spin relaxation rate can be essentially enhanced for graphene at
a certain type of substrate generating strong Rashba spin-orbit coupling, or in graphene with some heavy
adatoms\cite{castroneto09} and magnetic impurities\cite{kochan14} or for graphene with short-wavelength 
surface ripples.\cite{huertas06,dugaev11}

An additional mechanism of the spin relaxation can occur due to electron spin-flip scattering from
the magnetized graphene edge. Such an effect of the magnetization of zigzag edges has been
predicted by all first-principle as well as Hubbard-model calculations (for review, see 
Refs.~\onlinecite{katsnelson,yazyev10}), although experimental situation remains still 
not clear.  
Since the ordering of effectively one-dimensional system of magnetic moments at the 
edge is necessarily broken by fluctuations,\cite{yazyev08} the magnetized zigzag edge could be 
powerful source of the spin relaxation. However, as we demonstrate in this paper, this is not 
always the case because the scattering
probability of electrons at the zigzag edge is vanishingly small for gliding electrons.
Moreover, as we also show in this paper, in ballistic nanoribbons with magnetized zigzag edges, 
an anomalous diffusion of the spin density is possible, mostly due to electrons gliding along 
the ribbon edge.\cite{dugaev13}  

Our work is also motivated by the experiment,\cite{wojtaszek13} in which the enhanced spin 
relaxation time in hydrogenated graphene spin-valve devices has been found about 2~ns, so 
that the spin relaxation length was 7~$\mu $m.        

\section{Model}

We consider the graphene ribbon of width $L$ along the axis $y$, so that the graphene edges are
at $x=0$ and $x=L$ (see Fig.~1). 
We assume that the nanoribbon width $L$ is not so small to take into consideration
the size quantization, i.e., the wavelength of electrons at the Fermi level $\lambda _F$ is much smaller 
than $L$. It enables using the semiclassical Botzmann equation to describe the charge or spin transport.  
At the same time, we assume that $L$ can be of the order or smaller than the electron 
mean free path $\ell $. Therefore, in the following we consider in detail two main regimes: 
dirty nanoribbon when $\ell \ll L$, and ballistic nanoribbon with $\ell \gg L$.       

Let us consider first the case of a dirty nanoribbon.
The Boltzmann equation for the distribution function $f_{{\bf k}\sigma }({\bf r},t)$ of electrons with spin 
$\sigma =\uparrow, \downarrow $ is
\begin{eqnarray}
\label{1}
\frac{\partial g_{{\bf k}\sigma }}{\partial t}
+v_i\frac{\partial g_{{\bf k}\sigma }}{\partial r_i}
=\sum _{\bf k'} V_{\bf kk'}(g_{{\bf k'}\sigma }-g_{{\bf k}\sigma })
\hskip1cm
\nonumber \\
+[\delta (x)+\delta (x-L)]\, a_0 
\sum _{\bf k'} W_{\bf kk'}(g_{{\bf k'}\overline{\sigma }}
-g_{{\bf k}\sigma }),
\end{eqnarray}
where $V_{\bf kk'}$ is the probability of potential scattering of electrons in the bulk, 
$W_{\bf kk'}$ is the probablity of spin-flip scattering at the edges $x=0$
and $x=L$, $v_i=vk_i/k$, and we denote 
$f_{{\bf k}\sigma }({\bf r},t)=f_0(\varepsilon _{\bf k})+g_{{\bf k}\sigma }({\bf r},t)$,
where $f_0(\varepsilon _{\bf k})$ is the equilibrium Fermi-Dirac distribution function.
In Eq.~(1) we denoted by $\overline{\sigma }$ the opposite to $\sigma $ state, and
$a_0$ is the characteristic size of the transition region near the edge,\cite{ustinov80}
where the edge scatterers are located. In the following we take $a_0$ 
equal to the lattice constant of graphene.

\begin{figure}
\includegraphics[width=1.0\linewidth]{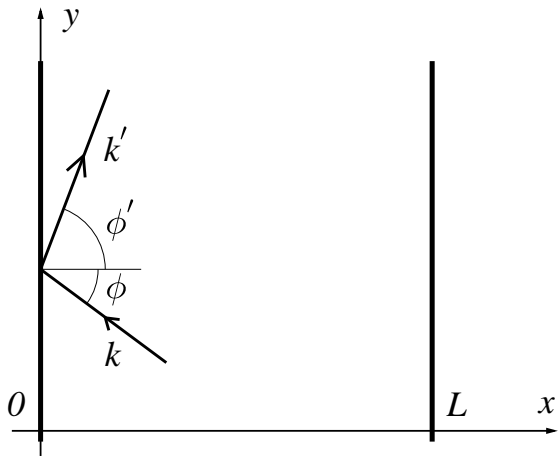}
\caption{Schematic presentation of the graphene ribbon along the axis $y$. Scattering of electrons from
the edges at $x=0$ and $x=L$ is associated with spin-flip transitions.}
\end{figure}

We do not include intervalley scattering to kinetic equation (1) assuming that the scatterers are not 
so short ranged. For the spin-flip edge scattering, it means that we assume that the correlation
length of magnetic fluctuations $R_c$ is large with respect to the lattice constant of
graphene, $R_c\gg a_0$. 
    
The kinetic equation of this form (1) with delta-functions in
the right hand part can be used to describe the dirty ribbon,
$\ell \ll L$. In the opposite ballistic case of $\ell \gg L$,
one should consider spin-flip scatterers as distributed within
the ribbon (see below).    

We can use Eq.~(1) to describe the diffusion of spin density
along the ribbon. Hence, we consider time evolution of the
spin distribution along the ribbon, defined as
\begin{eqnarray}
\label{2}
s_k(y,t)=\int _0^Ldx \int \frac{d\Omega _{\bf k}}{2\pi }
\left( g_{{\bf k}\uparrow }-g_{{\bf k}\downarrow }\right) ,
\end{eqnarray}
which includes averaging over orientation of the wavevector ${\bf k}$.
Here we denoted by $\Omega _{\bf k}$ the polar angle of the vector ${\bf k}$. 

If we choose the initial state with a smooth distribution $s_k(y,0)$
along $y$-axis, which does not depend on $x$, then 
\begin{eqnarray}
\label{3}
s_k(y,t)=L\int \frac{d\Omega _{\bf k}}{2\pi }
\left( g_{{\bf k}\uparrow }-g_{{\bf k}\downarrow }\right) .
\end{eqnarray}
So, in the following we concentrate on the one-dimensional spin diffusion along the 
nanoribbon, which is consistent 
with the experimental setup in Ref.~\onlinecite{wojtaszek13}.  

For distances along the ribbon large with respect to $\ell $
we can use the diffusive approximation. Thus, we substitute 
$g_{{\bf k}\sigma }(y,t)=p_{k\sigma }(y,t)+v_y \xi _{k\sigma }(y,t)$,
where the first term does not depend on direction of vector ${\bf k}$, and find
the diffusion equation for $p_{k\sigma }$
in the bulk
\begin{eqnarray}
\label{4}
\frac{\partial p_{k\sigma }}{\partial t}
-D\, \frac{\partial ^2p_{k\sigma }}{\partial y^2}=0,
\end{eqnarray}
and $\xi _{k\sigma }=-\tau \, (\partial p_{k\sigma }/\partial y)$,
where $D=v^2\tau $ is the one-dimensional diffusion coefficient, $\tau =\ell /v$,
and $p_{k\sigma }=\int \frac{d\Omega _{\bf k}}{2\pi }\, g_{{\bf k}\sigma }$.

The general solution of Eq.~(4) reads
\begin{eqnarray}
\label{5}
p_{k\sigma }(y,t)
=\int \frac{dq}{2\pi }\, B_{k\sigma }(q)\, e^{iqy-Dq^2t} .
\end{eqnarray}
It gives us
\begin{eqnarray}
\label{6}
g_{{\bf k}\sigma }(y,t)
=\int \frac{dq}{2\pi }\, B_{k\sigma }(q)\, 
\left( 1-i\tau vq\, \frac{k_y}{k}\right) e^{iqy-Dq^2t} .
\end{eqnarray}
where the coefficients $B_{k\sigma }(q)$ are related to the initial condition for the
distribution $g_{{\bf k}\sigma }(y,0)$.

Using Eqs.~(2)-(4) one can also write the diffusion equation for
spin density propagation in the bulk
\begin{eqnarray}
\label{7}
\frac{\partial s_k}{\partial t}
-D\, \frac{\partial ^2s_k}{\partial y^2}=0,
\end{eqnarray}
which has the general solution 
\begin{eqnarray}
\label{8}
s_k(y,t)=L\int \frac{dq}{2\pi } 
\left[ B_{k\uparrow }(q)-B_{k\downarrow }(q)\right] e^{iqy-Dq^2t} .
\end{eqnarray}
The next step is to establish boundary conditions for the kinetic equations.

\section{Boundary condition for the distribution function}

Now we assume that the edges of the graphene nanoribbon are magnetized.\cite{katsnelson,fujita96,yazyev10} 
It can be spontaneous magnetization as in the case of zigzag edge or due to magnetic impurities 
located at this edges.

The boundary condition at $x=0$ is related to the spin-flip scattering from magnetic moments
at the edge. It can be found by integrating Eq.~(1) over a smal region $a_0$ near the edge
\cite{ustinov80}
\begin{eqnarray}
\label{9}
|v_x|g_\sigma ^>(k_y,0)=|v_x|g_\sigma ^<(k_y,0)
+a_0\sum _{\bf k'} W_{\bf kk'}
\nonumber \\ \times
\left[ g_{\overline{\sigma }}^<(k'_y,0)-g_\sigma ^>(k_y,0)\right] .
\end{eqnarray}
%
We consider first the case of ribbon with $L\gg \ell $. 
Then the distribution of incoming electrons over the
angles is nearly homogeneous, $g_\sigma ^<(k_y,0)\simeq p_\sigma (0)$. 
Therefore, we can write Eq.~(9) as
\begin{eqnarray}
\label{10}
|v_x|\, g_\sigma ^>(k_y,0)=|v_x|\, p_\sigma (0)
+W_{\bf k}
\left[ p_{\overline{\sigma }}(0)-g_\sigma ^>(k_y,0)\right] ,\hskip0.3cm
\end{eqnarray}
where we denote $W_{\bf k}=a_0\sum _{\bf k'}W_{\bf kk'}$.

Similarly, we can obtain the corresponding boundary condition at $x=L$
\begin{eqnarray}
\label{11}
|v_x|\, g_\sigma ^<(k_y,L)=|v_x|\, p_\sigma (L)
+W_{\bf k}
\left[ p_{\overline{\sigma }}(L)-g_\sigma ^<(k_y,L)\right] ,\hskip0.3cm
\end{eqnarray}
Using Eqs.~(10) and (11) we find
\begin{eqnarray}
\label{12}
g_\sigma ^>(k_y,0)
=\frac{|v_x|}{|v_x|+W_{\bf k}}\, p_\sigma (0)
+\frac{W_{\bf k}}{|v_x|+W_{\bf k}}\, p_{\overline{\sigma }}(0)
\end{eqnarray}
and
\begin{eqnarray}
\label{13}
g_\sigma ^<(k_y,L)
=\frac{|v_x|}{|v_x|+W_{\bf k}}\, p_\sigma(L)
+\frac{W_{\bf k}}{|v_x|+W_{\bf k}}\, p_{\overline{\sigma }}(L).
\end{eqnarray}
The spin flow through the left edge (spin drain) is
\begin{eqnarray}
\label{14}
J_s(0)
=\int \frac{d\Omega _{\bf k}}{\pi }|v_x|\, 
\big[ p_{\uparrow }(0)-p_{\downarrow }(0)
-g_\uparrow ^>(k_y,0)\hskip0.5cm
\nonumber \\
+g_\downarrow ^>(k_y,0)\big] 
=2\big[ p_{\uparrow }(0)-p_{\downarrow }(0)\big]
\int \frac{d\Omega _{\bf k}}{\pi }\,
\frac{|v_x|\, W_{\bf k}}{|v_x|+W_{\bf k}}\, ,
\end{eqnarray}  
whereas the linear spin density in the ribbon is
(it does not depend on $x$) 
\begin{eqnarray}
\label{15}
s_k=L\big[ p_\uparrow (0)-p_\downarrow (0)\big] \, .
\end{eqnarray}
The outgoing spin current (spin drain) through the right edge is the same,
$J_s(L)=J_s(0)$.

Thus, we find the effective spin relaxation time
\begin{eqnarray}
\label{16}
\frac1{\tau _s}=\frac{2J_s(0)}{s_k}
=\frac4{\pi L}\int d\Omega _{\bf k}\,
\frac{|v_x|\, W_{\bf k}}{|v_x|+W_{\bf k}}\, .
\end{eqnarray} 
It corresponds to the relaxation of spin density in the nanoribbon.
Note that, in fact, the functions $p_{k\sigma }$ do not depend on $x$,
as discussed before.    

\section{Probability of spin-flip scattering at the edge}

As mentioned before, we assume that the main source of spin-flip scattering in the graphene nanoribbon 
is due to the magnetized graphene edge.\cite{fujita96,yazyev10,katsnelson} 
Recently, this problem has been very intensively studied by different methods, and therefore the 
possibility of edge magnetization is generally recognized by researchers.

We consider the $s$-$d$ exchange coupling Hamiltonian describing the coupling of electrons to 
the magnetic moments localized at the edge $x=0$
\begin{eqnarray}
\label{17}
H_{int}=W(x)\, \bsig \cdot {\bf m}(y), 
\end{eqnarray}  
where $W(x)$ is a short-range exchange potential related to the shape of 
electron wavefunction in a state localized at the edge. Similar spin-flip
scattering of electrons is at the other edge $x=L$.

Let us assume that the distrubution of the moments ${\bf m}(y)$ along the edge $x=0$ has the 
following form
\begin{eqnarray}
\label{18}
{\bf m}(y)={\bf m}_0+\delta {\bf m}(y),
\end{eqnarray}
where ${\bf m}_0$ is a constant part of magnetization 
(one can take it oriented along the quantization axis $z$) and
$\delta {\bf m}(y)$ is a random fluctuation part, such
that $\left< \delta {\bf m}(y)\right> =0$,
$\left< \delta m_\alpha (y)\, \delta m_\beta (y')\right> =
\gamma \, \delta _{\alpha\beta}\, e^{-\lambda |y-y'|}$ 
(here $\alpha ,\beta =x,y$), and $\lambda ^{-1}\equiv R_c$ is the 
correlation length of magnetic fluctuations.\cite{yazyev08} 
The constant $\gamma $ characterizes the amplitude of fluctuations,
$\gamma =\left< (\delta {\bf m})^2\right> $. 

The matrix element of spin-flip interaction (17) of electrons with the fluctuating 
magnetic moments
at the edge  (here $\psi _{\bf k}$ is spinor in the sublattice space of graphene)   
\begin{eqnarray}
\label{19}
w_{\bf kk'}=\int d^2{\bf r}\, \psi ^\dag _{\bf k}({\bf r})\,
W(x)\, [\delta m_x(y)-i\, \delta m_y(y)]\,
\psi _{\bf k'}({\bf r}), \hskip0.3cm 
\end{eqnarray}
and, after averaging over fluctuating moments, for the corresponding  spin-flip probability 
$W_{\bf kk'}=\frac{2\pi }{\hbar }\left< |w_{\bf kk'}|^2\right>
\delta (\varepsilon _k-\varepsilon _{k'})$ we obtain   
\begin{eqnarray}
\label{20}
W_{\bf kk'}=\frac{4\pi \gamma }{\hbar }
\int d^2{\bf r}\, d^2{\bf r'}\, e^{-\lambda |y-y'|} 
\psi ^\dag _{\bf k}({\bf r})\, W(x)\, \psi _{\bf k'}({\bf r})
\nonumber \\ \times
\psi ^\dag _{\bf k'}({\bf r'})\, W(x')\, \psi _{\bf k}({\bf r'})\, 
\delta (\varepsilon _{\bf k}-\varepsilon _{\bf k'}).\hskip0.3cm
\end{eqnarray}
To calculate the integrals in Eq.~(20) we need to know the wavefunctions
$\psi _{\bf k}({\bf r})$ near the graphene edge. It is well known that the choice
of these functions depends on the type of the edge, and can be described by
using different boundary conditions for the wave functions.  

\subsection{Berry-Mondragon boundary condition for the wavefunction at the edge}

Strictly speaking, in the case of Berry-Mondragon (BM) boundary we probably should not expect any spontaneous
magnetization of the graphene edge. Most probably, this effect is associated only with
the zigzag-shaped\cite{son06,yazyev08} and chiral\cite{yazyev11} edges.
Nevertheless, in the absence of any firm confirmation of these mostly theoretical suggestions,
we consider the effect of magnetization for different types of the boundaries.
Note that the magnetized boundary can be also achieved due to correlated magnetic atoms 
intentially inserted at the graphene edges.    

The Berry-Mondragon type of the boundary conditions\cite{berry87} relates the pseudo-spinor 
components of the wavefunction in graphene to those in the vacuum, described by a large gap 
in the graphene Hamiltonian. In this case the wave function near $x=0$ acquires to following 
form (here $\mathcal{L}$ is the ribbon length) 
\begin{eqnarray}
\label{21}
\psi _{\bf k}({\bf r})
=\frac{e^{i{\bf k}\cdot {\bf r}}}{\sqrt{2\mathcal{L}L}}
\left( \begin{array}{c} 1 \\ -i\end{array} \right) ,
\end{eqnarray}
and from Eq.~(20) we then obtain
\begin{eqnarray}
\label{22}
W^{(bm)}_{\bf kk'}=\frac{\pi \gamma W_0^2}{\hbar \mathcal{L}L^2}
\int dy\, e^{-\lambda |y|}e^{-i(k_y-k'_y)y}
\delta (\varepsilon _{\bf k}-\varepsilon _{\bf k'}),
\end{eqnarray}
where $W_0=\int dx\, e^{i(k-k')x}W(x)$ is a constant, provided the potential $W(x)$ is
short ranged.

After calculating the integral over $y$ in Eq.~(22) we come to
\begin{eqnarray}
\label{23}
W^{(bm)}_{\bf kk'}=\frac{2\pi \gamma W_0^2\lambda \, 
\delta (\varepsilon _{\bf k}-\varepsilon _{\bf k'})}
{\hbar \mathcal{L}L^2\big[ \lambda ^2+(k_y-k'_y)^2\big] }\, .
\end{eqnarray}
Then, integrating (23) over ${\bf k'}$ we find the function $W_{\bf k}$, which 
determines the spin relaxation time in Eq.~(16)
\begin{eqnarray}
\label{24}
W^{(bm)}_{\bf k}
=\frac{2\gamma W_0^2a_0\zeta }{\hbar ^2Lv}
\int _{-\pi /2}^{\pi /2}
\frac{d\phi '}{1+\zeta ^2(\sin \phi -\sin \phi ')^2}\, ,
\end{eqnarray}
where we denote the parameter $\zeta =kR_c$ and introduce the angles $\phi ,\phi '$ of,
respectively, incoming and outgoing electrons by $k_y=k\sin \phi $ and
$k'_y=k\sin \phi '$ (see Fig.~1). Of course, the value $k=k_F$ is relevant for the
electronic transport.

\begin{figure}
\hspace*{-1cm}
\includegraphics[width=1.1\linewidth]{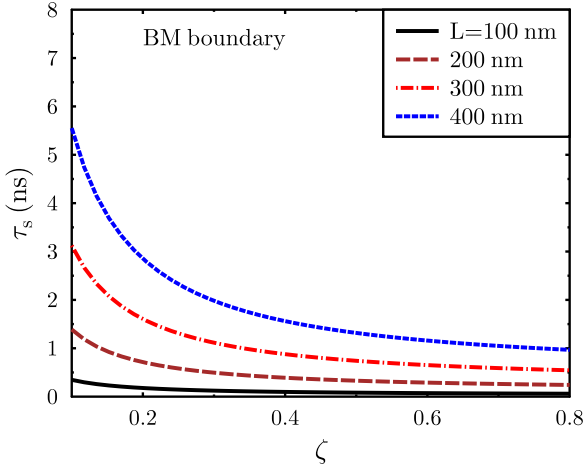}
\caption{Dependence of effective spin relaxation time on the parameter $\zeta $ for
different values of the ribbon width. The case of BM boundary.}
\end{figure}

\begin{figure}
\hspace*{-1cm}
\includegraphics[width=1.1\linewidth]{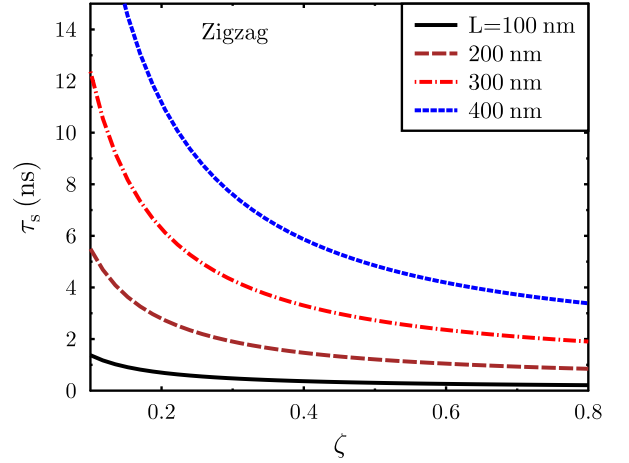}
\caption{The same as in Fig.~2 for the case of zigzag boundary.}
\end{figure}

The results of numerical calculations with (16) and (24) are presented in Fig.~2, where
we used the following parameters: $v=10^8$~cm/s, $\gamma =1$, $a_o=5\times 10^{-8}$~cm,
$W_0/a_0=0.2$~eV.

\subsection{Zigzag boundary}

In the case of zigzag boundary of graphene, the possibility of edge magnetization has been 
established in several works (see, e.g., the review article of Yazyev\cite{yazyev10} and Chap.~12 of 
Ref.~\onlinecite{katsnelson}). 

For the zigzag boundary we can use the wavefunction near the edge $x=0$ in the following
form\cite{brey06}
\begin{eqnarray}
\label{25}
\psi _{\bf k}({\bf r})
\simeq \frac{e^{ik_yy}}{\sqrt{\mathcal{L}L}} 
\left( \begin{array}{c} 
\sin k_xx
\\
 -\frac{ik_x}{k}\cos k_xx+\frac{ik_y}{k}\sin k_xx
\end{array} \right) .
\end{eqnarray}
This type of the boundary condition corresponds to requirement of zero value for one of the 
spinor components, and nonzero value for another one.   

Substituting (25) in Eq.~(20) we obtain
\begin{eqnarray}
\label{26}
W^{(z)}_{\bf kk'}=\frac{\pi \gamma W_0^2k_x^2k'^2_x}{\hbar \mathcal{L}L^2k^4}
\int dy\, e^{-\lambda |y|}e^{-i(k_y-k'_y)y}
\delta (\varepsilon _{\bf k}-\varepsilon _{\bf k'}),\hskip0.3cm
\end{eqnarray}
and, correspondingly, we get
\begin{eqnarray}
\label{27}
W^{(z)}_{\bf kk'}=\frac{2\pi \gamma W_0^2k_x^2k'^2_x\lambda \, 
\delta (\varepsilon _{\bf k}-\varepsilon _{\bf k'})}
{\hbar \mathcal{L}L^2k^4\big[ \lambda ^2+(k_y-k'_y)^2\big] }\; .
\end{eqnarray}
Now we can also calculate the function $W^{(z)}_{\bf k}$
\begin{eqnarray}
\label{28}
W^{(z)}_{\bf k}
=\frac{2\gamma W_0^2a_0\zeta \cos ^2\phi }{\hbar ^2Lv}
\int _{-\pi /2}^{\pi /2}
\frac{\cos ^2\phi '\, d\phi '}{1+\zeta ^2(\sin \phi -\sin \phi ')^2}\, .\hskip0.3cm 
\end{eqnarray}
Then using Eqs.~(16) and (28) we calculated numerically the spin relaxation time for the case of 
nanoribbon with zigzag edges in the regime of $\ell \ll L$. The dependence of
$\tau _s(\zeta )$ for the same parameters as in Fig.~1 are presented in Fig.~3.  

\section{Anomalous spin diffusion in the ballistic nanoribbon} 

In the ballistic regime of $L\ll \ell $ we can completely neglect the dependence 
of the distribution function on $x$ in Eq.~(1), and consider the spin-flip scattering 
as scattering within the bulk. It should be stressed that this condition of $L\ll \ell $
is, in fact, limiting only the value of $L$, whereas the mean free path is still finite
(and in principle can be not so large) which justifies using the diffusion approach 
(see below).   

Then, by using the diffusive approximation for the $y$-dependence of spin-resolved 
distribution, one can obtain  
\begin{eqnarray}
\label{29}
\frac{\partial p_{k\sigma }}{\partial t}-D\frac{\partial ^2p_{k\sigma }}{\partial y^2}
=\sum _{\bf k'} W_{\bf kk'}(p_{k\overline{\sigma }}-p_{k\sigma }),
\end{eqnarray}
For the case of BM edge we substitute (23) into Eq.~(29)
\begin{eqnarray}
\label{30}
\frac{\partial p_{k\sigma }}{\partial t}-D\frac{\partial ^2p_{k\sigma }}{\partial y^2}
=\frac{2\pi \gamma W_0^2\lambda (p_{k\overline{\sigma }}-p_{k\sigma })}{\hbar \mathcal{L}L^2}
\nonumber \\ \times
\sum _{\bf k'} \frac{\delta (\varepsilon _{\bf k}-\varepsilon _{\bf k'})}
{ \lambda ^2+(k_y-k'_y)^2}
\end{eqnarray}
and find the diffusion equation for spin density
\begin{eqnarray}
\label{31}
\frac{\partial s_k}{\partial t}-D\frac{\partial ^2s_k}{\partial y^2}=-\frac{s_k}{\tau _s}\, ,
\end{eqnarray}
where
\begin{eqnarray}
\label{32}
\frac1{\tau _s}
=\frac{2\gamma W_0^2\zeta }{\hbar ^2Lv}
\int _{-\pi /2}^{\pi /2}
\frac{d\phi '}{1+\zeta ^2(\sin \phi -\sin \phi ')^2}\, .
\end{eqnarray}
The dependence of $\tau _s/\tau _0$ on the angle $\phi $ is presented in Fig.~4, 
where we denoted
$\tau _0^{-1}=2\pi \gamma W_0^2\zeta /\hbar ^2Lv$.
As we see, the spin relaxation time is anisotropic (i.e., depending on the angle $\phi $
of incoming electrons) but this anisotropy is relatively weak.
It justifies the diffusive equation (31) for ballistic nanoribbon with BM boundaries.

\begin{figure}
\hspace*{-1cm}
\includegraphics[width=1.1\linewidth]{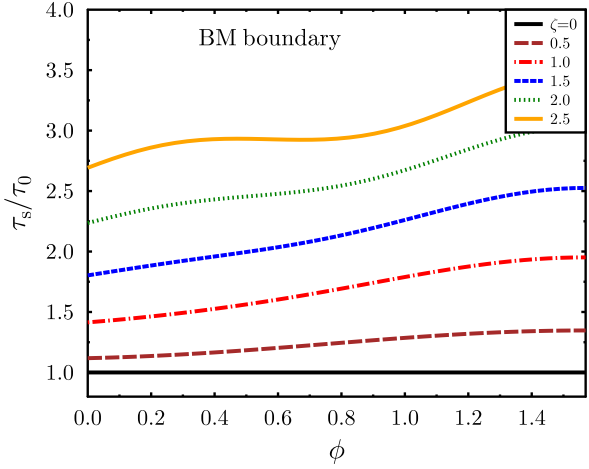}
\caption{The dependence of spin relaxation time on the angle $\phi $ 
in the ballistic regime with BM boundaries.}
\end{figure}

\begin{figure}
\hspace*{-1cm}
\includegraphics[width=1.1\linewidth]{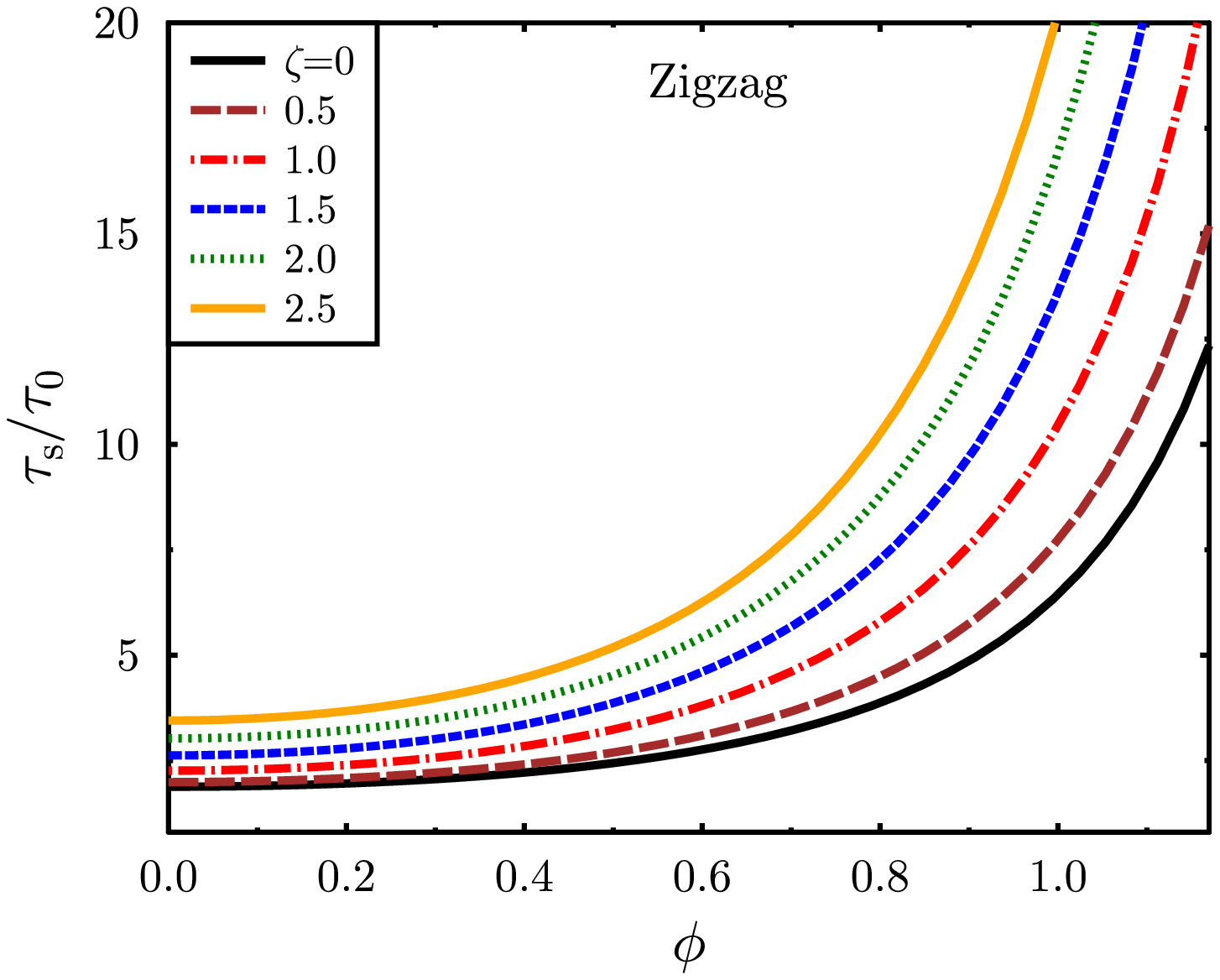}
\caption{The dependence of spin relaxation time on the angle $\phi $ 
in the ballistic regime with zigzag boundaries.}
\end{figure}

If we try to do the same for the ballistic ribbon with zigzag edges, then after using (26) we come to
the following result
\begin{eqnarray}
\label{33}
\frac1{\tau _s}
=\frac{2\gamma W_0^2\zeta \cos ^2\phi }{\hbar ^2Lv}
\int _{-\pi /2}^{\pi /2}
\frac{\cos ^2 \phi  '\, d\phi '}{1+\zeta ^2(\sin \phi -\sin \phi ')^2}
\end{eqnarray}
The corresponding dependence of $\tau _s$ on the angle $\phi $ is presented in Fig.~5 .
In this case the anisotropy of spin relaxation time is very strong. 

Strictly speaking, the diffusion equation (45) cannot be used to describe the spin propagation 
for the ribbon with zigzag edges because gliding electrons can propagate at a long distance
without changing their spin. Correspondingly, we anticipate that the ballistic transport of charge 
in such a ribbon is accompanied by the ballistic spin transport.       

\subsection*{Anomalous spin diffusion in the zigzag nanoribbon}

In the case of ballistic zigzag nanoribbon, by using the Boltzmann equation for $g_{{\bf k}\sigma}$ we 
can find the kinetic equation for $s_{\bf k}\equiv g_{{\bf k}\uparrow }-g_{{\bf k}\downarrow }$
in the following form 
\begin{eqnarray}
\label{34}
 \frac{\partial s(\phi )}{\partial t}+v\sin \phi \, \frac{\partial s(\phi) }{\partial y}
=-\frac{\cos ^2\phi }{\tau _0}\hskip1cm
\nonumber \\ \times
\int _{-\pi /2}^{\pi /2}
\frac{\cos ^2\phi '\, [s(\phi )+s(\phi ')]\, d\phi '}{1+\zeta ^2(\sin \phi -\sin \phi ')^2} \, ,
\end{eqnarray}
where $s(\phi )$ explicitly shows the dependence of spin distribution on the orientation of vector 
${\bf k}$.  

Let us consider the limiting cases of short and long correlation lengths $R_c$ in comparison
with electron wavelength $\lambda _F$.

\subsubsection{Large magnetic correlation length, $\zeta \gg 1$} 

In the case of $\zeta \gg 1$ (the correlation length is much larger than the electron wavelength at the
Fermi level $\lambda _F$), the main contribution in intergral (34) is from $\phi '\simeq \phi $. 
Then we can simplify Eq.~(34) to
\begin{eqnarray}
\label{35}
 \frac{\partial s(\phi )}{\partial t}+v\sin \phi \, \frac{\partial s(\phi) }{\partial y}
=-\frac{2C\cos ^3\phi \, s(\phi )}{\tau _0}\, ,
\end{eqnarray}
where we denote
\begin{eqnarray}
\label{36}
C=\frac1{\cos \phi } \int _{-\pi /2}^{\pi /2}
\frac{\cos ^2\phi '\, d\phi '}{1+\zeta ^2(\sin \phi -\sin \phi ')^2} 
\simeq \frac{\pi }{\zeta } \, .
\end{eqnarray}
After Fourier transformation over $y$, Eq.~(35) reads
\begin{eqnarray}
\label{37}
 \frac{\partial s_q(\phi )}{\partial t}+ivq\sin \phi \, s_q(\phi) 
=-\frac{2\pi \cos ^3\phi \, s_q(\phi )}{\tau _0\zeta }\, .
\end{eqnarray} 
This equation has the solution 
\begin{eqnarray}
\label{38}
s(\phi )=\int \frac{dq}{2\pi }R_q
\exp \left( iqy-ivqt\sin \phi -\frac{2\pi t\cos ^3\phi }{\tau _0\zeta }\right) ,\hskip0.3cm
\end{eqnarray}
where the arbitrary function $R_q$ should be related to the initial condition. 
We choose $R_q=1$, which corresponds to the $\delta $-function spin distribution at $t=0$.

Now we integrate (38) over angle $\phi $. It gives us the spin distribution at the moment $t$   
\begin{eqnarray}
\label{39}
s(t)=\frac1{\sqrt{v^2t^2-y^2}}\, \exp \left[ -\frac{2\pi t}{\tau _0\zeta }
\left( 1-\frac{y^2}{v^2t^2}\right) ^{3/2}\right] ,
\end{eqnarray}   
where $|y|<vt$.
 
Then we find
\begin{eqnarray}
\label{40}
\left< y^2(t)\right> 
\equiv \int _{-vt}^{vt} dy\, y^2\, s(t)
=v^2t^2\int _{-1}^1 \frac{z^2\, dz}{\sqrt{1-z^2}}\, 
\nonumber \\ \times
\exp \left[ -\frac{2\pi t}{\tau _0\zeta }\left( 1-z^2\right) ^{3/2}\right] ,
\end{eqnarray}
where $z=y/vt$.
The dependence on $t$ of the integral in Eq.~(40) describes deviation from pure ballistics. 
As follows from Eq.~(40), for $t\ll \tau _0\zeta $ the spin transport is ballistic.

At large times, $t\gg \tau _0\zeta $, we obtain from (40)
\begin{eqnarray}
\label{41}
\left< y^2(t)\right> 
\simeq \frac{2v^2t^2}3 \int _0^1 \frac{x^{-2/3}dx}{\sqrt{1-x^{2/3}}}\, e^{-2\pi tx/\tau _0\zeta } \hskip0.3cm
\\
\simeq \frac{2v^2(\tau _0\zeta )^{1/3}t^{5/3}\Gamma (\frac43 )}{(2\pi )^{1/3}} \hskip0.3cm
\end{eqnarray}
It corresponds to anomalous spin diffusion with characteristic diffusion length  
$l_{d} \equiv \sqrt{\left< y^2(t)\right> }\sim t^{5/6}$.

\subsubsection{Small correlation length, $\zeta \ll 1$}

Now we consider the case of small magnetic correlation length, i.e., $\zeta \ll 1$.
In this case we obtain from (34) (note that in the case of small $\zeta $, the characteristic 
relaxation time $\tau _0$ is large since $\tau _0\sim \zeta ^{-1}$)
\begin{eqnarray}
\label{43}
\frac{\partial s(\phi )}{\partial t}+v\sin \phi \, \frac{\partial s(\phi) }{\partial y}
=-\frac{\pi \cos ^2\phi \, s(\phi )}{2\tau _0}\hskip0.5cm
\nonumber \\ 
-\frac{\cos ^2\phi }{\tau _0}
\int _{-\pi /2}^{\pi /2}
\cos ^2\phi '\, s(\phi ')\, d\phi '.
\end{eqnarray}
The solution of Eq.~(43)  is
\begin{eqnarray}
\label{44}
s(\phi ,t,y)
=\delta (y-vt\sin \phi )\, e^{-\pi t\cos ^2\phi /2\tau _0}-\frac{\cos ^2\phi }{\tau _0}\hskip0.5cm
\nonumber \\ \times
\int _{y-vt}^{y+vt} dy'  
\int _0^{t-|y-y'|/v} dt'\, \delta [y-y'-v(t-t')\sin \phi ]
\nonumber \\ \times \,
e^{-\pi (t-t')\cos ^2\phi /2\tau _0} A(t',y') , \hskip0.5cm
\end{eqnarray}
where we denoted
\begin{eqnarray}
\label{45}
A(t,y)=\int _{-\pi /2}^{\pi /2}\cos ^2\phi \; s(\phi ,t,y)\, d\phi .
\end{eqnarray}
After multiplying (44) by $\cos ^2\phi $ and integrating over $\phi $ one can obtain the 
integral equation for the function $A(t,y)$
\begin{eqnarray}
\label{46}
A(t,y)=\frac{\sqrt{v^2t^2-y^2}}{v^2t^2}\, e^{-\pi t(1-y^2/v^2t^2)/2\tau _0}
\theta (vt-|y|) \hskip0.5cm
\nonumber \\ 
-\frac{1}{\tau _0} \int _{y-vt}^{y+vt}dy' 
\int _0^{t-|y-y'|/v} dt'\,
\frac{[v^2(t-t')^2-(y-y')^2]^{3/2} }{v^4(t-t')^4}
\nonumber \\ \times \,
e^{-\pi (t-t')[1-(y-y')^2/v^2(t-t')^2]/2\tau _0}A(t',y') .\hskip0.3cm
\end{eqnarray}

The integrated over angles spin distribution can be found from Eq.~(44)
\begin{eqnarray}
\label{47}
s(t,y)=\frac{e^{-\pi t(1-y^2/v^2t^2)/2\tau _0}}{\sqrt{v^2t^2-y^2}} \,
\theta (vt-|y|) \hskip1cm
\nonumber \\ 
-\frac{1}{\tau _0} \int _{y-vt}^{y+vt} dy' \int _0 dt'
\frac{\sqrt{v^2(t-t')^2-(y-y')^2}}{v^2(t-t')^2} 
\nonumber \\ \times \,
e^{-\pi t[1-(y-y')^2/v^2(t-t')^2]/2\tau _1}A(t',y') .
\end{eqnarray}
Using spin distribution (47) we can also find the mean square distance, at which
the spin density propagates in time $t$
\begin{eqnarray}
\label{48}
\left< y^2(t)\right> 
=v^2t^2\int _{-1}^1 \frac{z^2e^{-\pi t(1-z^2)/2\tau _0}dz}{\sqrt{1-z^2}} 
-\frac{v^3t^4}{\tau _1}  
\int _{-\infty }^{\infty } z^2 dz 
\nonumber \\ \times
\int _{z-1}^{z+1} dz'
\int _0^{1-|z-z'|} d\xi \; \frac{\sqrt{(1-\xi )^2-(z-z')^2}}{(1-\xi )^2}
\nonumber \\ \times
 e^{-\pi t(1-\xi )[1-(z-z')^2]/2\tau _0} A(t\xi ,z') ,\hskip0.5cm
\end{eqnarray}
where we denoteed $z=y/vt$, $z'=y'/vt$, and $\xi = t'/t$.
If $t\ll \tau _0$ then the spin propagation is pure ballistic at a large distance, $l\simeq v\tau _0$.

If $t\gg \tau _0$, the main contribution to the first integral in (48) is from $z\simeq 1$, 
and in the second integral from $\xi \sim 1$ and $|z-z'|\ll 1$. Then we obtain from (48)
\begin{eqnarray}
\label{49}
\left< y^2(t)\right> 
\simeq 2^{-1/2}v^2t^{3/2}\tau _0^{1/2}  
-\frac{v^3t^4}{\tau _0}  \int _{-\infty }^{\infty } z^2 dz \int _{0}^{1} dr 
\int _0^{1-r} d\xi 
\nonumber \\ \times
\frac{[A(t\xi ,z+r)+A(t\xi ,z-r)]}{1-\xi }\;
 e^{-\pi t(1-\xi )(1-r)/\tau _0} .\hskip0.5cm
\end{eqnarray}
The function $A(t,y)$ is calculated in Appendix. 
Since $A(t,z)$ exponentially decays at $t\gg \tau _0$,  the  
main contribution in the last integral of (49) is from the vicinity of $r\sim 1$ 
and $\xi \ll 1$. Correspondingly, we get
\begin{eqnarray}
\label{50}
\left< y^2(t)\right> 
\simeq 2^{-1/2}v^2t^{3/2}\tau _0^{1/2}  
-\frac{v^3t^4}{\tau _0}  \int _{-\infty }^{\infty } z^2 dz 
\nonumber \\ \times
 \int _{0}^{1} dx \int _0^x d\xi \; A(t\xi ,z\pm 1)\;
 e^{-\pi tx/\tau _0} .\hskip0.5cm
\end{eqnarray}
Here the main contribution to the integral comes from $x\sim \xi \sim \tau _0/t\ll 1$.
Then using (A10) we finally obtain
\begin{eqnarray}
\label{51}
\left< y^2(t)\right> \simeq \left( \frac1{\sqrt{2}}-\frac2{\pi ^2}\right) 
v^2t^{3/2}\tau _0^{1/2} , 
\end{eqnarray}
which leads to the anomalous spin diffusion law with the diffusion length $l_{d}\sim t^{3/4}$.
 
\section{Conclusion}

We studied the effect of spin-flip scattering of electrons from the magnetized edges
of graphene nanoribbons. The essential point of our model is an assumption of strong
fluctuations of the magnetic moments at the graphene edge, which have been established earlier 
in Ref.~\onlinecite{yazyev08}. The spin-flip scattering of this type, which is relevant for spin 
relaxation in graphene nanoribbons, can be strongly suppressed for electron incoming under
small angles to the edge. We found that this effect is especially strong for the zigzag 
boundary of graphene.
For such gliding electrons the spin is nearly conserved. As a result, there is a possibility 
of anomalous spin diffusion along the graphene nanoribbon.    
     
For the estimation of parameters we can use the relation for conductivity in graphene
$\sigma =e^2k_F\ell /\pi \hbar $, which gives us for the mobility $\mu =e\tau v/\hbar k_F$. 
For the carrier density in graphene we use $n=k_F^2/\pi $.
We can assume the mean free path of electrons in the bulk $\ell =3\times 10^{-4}$~cm and 
$n=10^{12}$~cm$^{-2}$. It gives us the bulk relaxation time $\tau =\ell /v=3\times 10^{-12}$~s, 
$k_F\simeq 1.8\times 10^6$~cm$^{-1}$ and $\mu \simeq 2.5\times 10^5$~cm$^2$/V$\cdot $s,
which can be achieved in graphene.\cite{mayorov11}
This gives us the estimation of characteristic width for the ribbon, that is, the ballistic
case corresponds to $L\ll 10^{-4}$~cm. 

\begin{figure}
\hspace*{-1cm}
\includegraphics[width=1.1\linewidth]{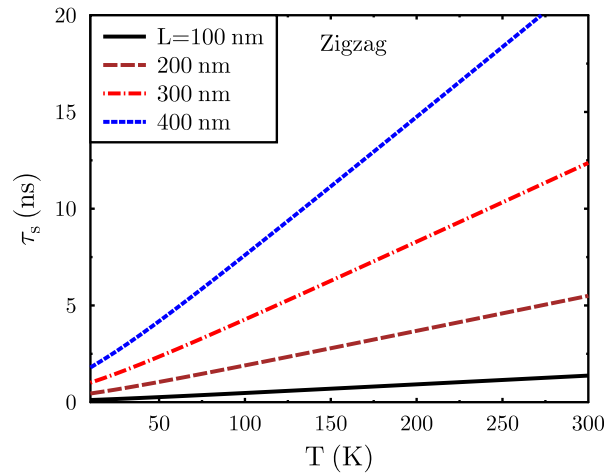}
\caption{Dependence of spin relaxation time on temperature for
different values of the ribbon width. The case of zigzag boundary, and we take 
$k_F=1\times 10^6$~cm$^{-1}$.}
\end{figure}

The correlation length $R_c$ has been calculated in Ref.~\onlinecite{yazyev08} as a function 
of temperature. According to this estimation in can vary from 1000~nm at very low temperatures
to about 1~nm at the room temperature. 
Thus, for the presented above parameters of graphene
(i.e., not too small density of electrons) we can expect the value of parameter 
$\zeta =k_FR_c\ll 1$. In our approach it corresponds to the case of small correlation length.

It should be noted that the temperature dependence of correlation length $R_c$  
determines the temperature dependence of spin relaxation in our model. 
In temperature range of 10 to 100~K, one can approximate it \cite{yazyev08} by 
$R_c\simeq 100/k_BT$~(nm). Then according to Eqs.~(16),(28) the relaxation time grows 
with the temperature. Figure~6 demonstrates that this dependence is almost linear. 
At high temperatures, when the correlation length is much smaller than the electron wave length,
electrons do not feel fluctuating spins because the magnetic disorder is effectively
averaged over the wave length.     
%

\section*{Acknowledgements}

We thank E. Sherman for discussions.
The work of VKD is supported by the National Science Center in Poland by the Grant 
No.~DEC-2012/06/M/ST3/00042. 
MIK acknowledges funding from the European Union Seventh Framework Programme under 
grant agreement No.~604391 Graphene Flagship and from ERC Advanced Grant No.~338957 FEMTO/NANO.

\begin{appendix}

\section{Calculation of the function $A(t,y)$}

Using the Fourier transformation of Eq.~(43)  
\begin{eqnarray}
\label{a1}
\left( -i\omega +ivq\sin \phi +\frac{\pi \cos ^2\phi }{2\tau _0}\right) s_{\omega q}
=-\frac{\cos ^2\phi \, A_{\omega q}}{\tau _0}
\end{eqnarray}
we find
\begin{eqnarray}
\label{a2}
s_{\omega q}
=-\frac{\cos ^2\phi \, A_{\omega q}}
{-i\omega \tau _0+ivq\tau _0\sin \phi +\frac{\pi }2 \cos ^2\phi }\, .
\end{eqnarray}
Then after multilpying (A2) by $\cos ^2\phi $ and integrating over angle $\phi $ we come to
\begin{eqnarray}
\label{a3}
A_{\omega q}\left( 
1+\int _{-\pi /2}^{\pi /2}\frac{\cos ^4\phi \, d\phi }
{-i\omega \tau _0+ivq\tau _0\sin \phi +\frac{\pi}2 \cos ^2\phi }\right) =0.\hskip0.4cm
\end{eqnarray}
From (A3) the condition of nonzero $A_{q\omega }$ gives us the equation
\begin{eqnarray}
\label{a4}
1+\int _{-\pi /2}^{\pi /2}\frac{\cos ^4\phi \, d\phi }
{-i\omega \tau _0+ivq\tau _0\sin \phi +\frac{\pi }2 \cos ^2\phi }=0
\end{eqnarray}
determining the dependence $\omega (q)$.

Thus, the function $A(t,y)$ can be presented in the form
\begin{eqnarray}
\label{a5}
A(t,y)=\int \frac{dq}{2\pi }\; e^{iqy-i\omega (q)\, t}\, ,
\end{eqnarray}
where we choose $A_{\omega q}=1$, which corresponds to the assumed initial spin distribution 
$s(t=0,y)$.

Let us look for the solution of Eq.~(A4) in the form
\begin{eqnarray}
\label{a6}
\omega (q)=-ix(q)/\tau _0
\end{eqnarray}
with $x(q)$ real. Then for $t\gg \tau _0$ the integral over $q$ in (A5) is mostly determined by 
$q$ near the minimum of dependence $x(q)$. 

Substituting (A6) to (A4) we obtain
\begin{eqnarray}
\label{a7}
\int _{-\pi /2}^{\pi /2}\frac{\cos ^4\phi \, d\phi }
{x-i\kappa \sin \phi -\frac{\pi }2 \cos ^2\phi }=1
\end{eqnarray}
where we denoted $\kappa =vq\tau _0$.

\begin{figure}
\hspace*{-1cm}
\includegraphics[width=1.1\linewidth]{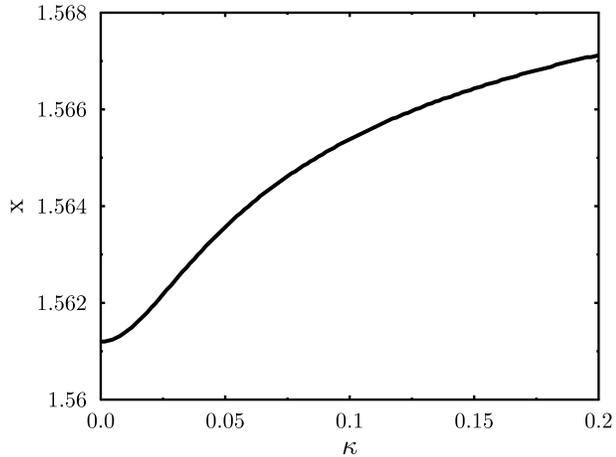}
\caption{The dependence $x(\kappa )$ calculated numerically from Eq.~(A7).}
\end{figure}

The dependence $x(\kappa )$ calculated numerically from (A7) is shown in Fig.~7. 
It can be interpreted as the diffusion mode
of a partial spin density related to electrons moving in transversal direction. Such electrons
are strongly scattered from the edges, and therefore such mode is exponentially decaying (it has a gap).     
In the vicinity of $\kappa =0$, it can be approximated by $x(\kappa )\simeq \beta +\alpha \kappa ^2$,
with $\beta \simeq 1.561$.
Indeed, here the characteristic values are $y\sim vt$, $q\sim 1/y\sim 1/vt$, so that $\kappa \sim \tau _0/t\ll 1$,
which corresponds to very close vicinity of the minimum in Fig.~7. 

Substituting this approximation to (A5) we find
\begin{eqnarray}
\label{a8}
A(t,y)=\frac{e^{-\beta t/\tau _1}}{vt^{1/2}\tau _1^{1/2}}\; \psi (\tilde{y}),
\end{eqnarray}
where $\tilde{q}=v\sqrt{t\tau _0}\, q$, $\tilde{y}=y/v\sqrt{t\tau _0}$ and
\begin{eqnarray}
\label{b11}
\psi (\tilde{y})=\frac{e^{-\tilde{y}^2/4\alpha }}{2\sqrt{\pi \alpha }}
\end{eqnarray} 
is a universal function which does not depend on any parameters.

Substituting (A9) to the integral for $A(\tau _0,z)$ 
and using relation $z=y/vt=(\tau _0/t)^{1/2}\tilde{y}$, we obtain
\begin{eqnarray}
\label{a9}
\int z^2dz\, A(\tau _0,z\pm 1)
\simeq \frac2{v\sqrt{t\tau _0}}\, .
\end{eqnarray}

\end{appendix}

\end{document}